# Artificial intelligence for context-aware visual change detection in software test automation

Milad Moradi[*]
AI Research Lab, Tricentis, Vienna, Austria
m.moradi-vastegani@tricentis.com
ORCID: 0000-0002-9724-0339

Ke Yan
AI Research Lab, Tricentis, Sydney, Australia
k.yan@tricentis.com

David Colwell
AI Research Lab, Tricentis, Sydney, Australia
d.colwell@tricentis.com

Rhona Asgari
AI Research Lab, Tricentis, Vienna, Austria
r.asgari@tricentis.com

[*] Corresponding author. **Postal address:** Tricentis GmbH, Leonard-Bernstein-Straße 10, 1220 Vienna, Austria.

**Abstract**

Automated software testing is integral to the software development process, streamlining workflows and ensuring product reliability. Visual testing, particularly for user interface (UI) and user experience (UX) validation, plays a vital role in maintaining software quality. However, conventional techniques such as pixel-wise comparison and region-based visual change detection often fail to capture contextual similarities, subtle variations, and spatial relationships between UI elements. In this paper, we propose a novel graph-based approach for context-aware visual change detection in software test automation. Our method leverages a machine learning model (YOLOv5) to detect UI controls from software screenshots and constructs a graph that models their contextual and spatial relationships. This graph structure is then used to identify correspondences between UI elements across software versions and to detect meaningful changes. The proposed method incorporates a recursive similarity computation that combines structural, visual, and textual cues, offering a robust and holistic model of UI changes. We evaluate our approach on a curated dataset of real-world software screenshots and demonstrate that it reliably detects both simple and complex UI changes. Our method significantly outperforms pixel-wise and region-based baselines, especially in scenarios requiring contextual understanding. We also discuss current limitations related to dataset diversity, baseline complexity, and model generalization, and outline planned future improvements. Overall, our work advances the state of the art in visual change detection and provides a practical solution for enhancing the reliability and maintainability of evolving software interfaces.

**Keywords:** Software test automation, Artificial intelligence, Machine learning, Computer vision, Graph-based visual change detection

# 1. Introduction

Quality assurance in software development plays a pivotal role in ensuring the reliability, functionality, and overall excellence of software products [1, 2]. As the complexity of software systems continues to escalate, the significance of robust quality assurance practices becomes increasingly important. Quality assurance is a comprehensive process that spans the entire software development life cycle, from initial design and coding to testing and deployment. It involves systematic and meticulous testing, verification, and validation processes aimed at identifying and rectifying defects, bugs, and potential issues before a software product reaches end-users [3-5].

Software test automation is a cornerstone of modern software development, offering enhanced efficiency, repeatability, and scalability [6]. In the fast-paced landscape of continuous integration and deployment, test automation helps ensure application stability and reduces regression risks [7]. A particularly challenging aspect of automated testing is visual change detection, which involves identifying unintended visual alterations in user interfaces (UIs) across different versions or platforms of an application. Visual regressions can significantly impact user experience, yet they are often missed by traditional automated tests.

Conventional visual change detection techniques, such as pixel-wise comparison or region-based matching, are widely used due to their simplicity, but they fail to capture semantic meaning, contextual relationships, or layout variability across platforms and screen sizes. These methods are highly sensitive to benign changes such as minor alignment shifts or colour variations, leading to false positives and missed true regressions in more complex UIs. Moreover, they lack robustness when dealing with cross-platform interfaces where element locations and hierarchies may differ. As such, existing solutions fall short in real-world UI testing scenarios, particularly in mobile-responsive applications and layout-driven designs [8-14].

To address these challenges, we introduce a novel method for visual change detection in software test automation, leveraging Artificial Intelligence (AI), Machine Learning (ML), and graph-based modelling to achieve context-aware analysis. The core of our approach lies in:

1) Employing a deep learning object detection model, i.e. You Only Look Once (YOLO) [15] to detect and label UI controls within screenshots;
2) Constructing a graph representation of the UI layout, where nodes denote detected controls and edges represent spatial relationships;
3) Introducing a recursive graph similarity algorithm that incorporates visual, textual, and neighborhood-based similarity to find corresponding elements across screenshots;

4) Using unmatched nodes to detect and localize meaningful UI changes via heatmaps.

This method enables a semantic, layout-aware comparison between software versions, capable of identifying both subtle and structural visual changes across desktop and mobile platforms.

We conducted comprehensive experiments across three datasets simulating realistic test scenarios: (1) UI changes on desktop browsers; (2) altered layouts due to window resizing; and (3) cross-platform changes between desktop and mobile views. Our results show that the proposed method consistently outperforms traditional pixel-wise and region-based baselines, especially in complex or cross-platform settings. These findings highlight the robustness and practical value of our approach for scalable and intelligent UI regression testing.

The rest of this paper is organized as follows. In Section 2, we give an overview of related work in software test automation and AI-driven change detection in images. A detailed description of our graph-based visual change detection method is given in Section 3. We then present and discuss the experimental results in Section 4. Finally, concluding remarks are discussed and future lines of work are drawn in Section 5.

## 2. Related work

Software testing is a critical and integral component of the software development life cycle, ensuring the quality, reliability, and functionality of software products [16]. Automated software testing enhances this process by increasing efficiency, consistency, and scalability [17, 18], and it has been widely adopted in modern development pipelines using techniques such as unit, integration, and regression testing [19-21]. Visual testing, which ensures that UI and UX components meet design expectations across platforms, plays a central role in detecting visual regressions often missed by traditional functional tests [8-11]. However, traditional methods for visual change detection, such as pixel-wise comparison or region-based matching, are highly sensitive to benign visual differences and lack semantic understanding or layout robustness [12, 13].

In response, AI and machine learning have been adopted to provide more context-aware and intelligent testing solutions. Prior studies have applied AI to test case generation, prioritization, self-healing, and UI interaction modeling [22-26]. Within the domain of visual change detection, object detection techniques have proven particularly effective in extracting high-level UI representations. Our work builds on this by employing a CNN-based model (YOLO) to detect UI controls and represent UI layouts as graphs, enabling semantic and structural comparison.

However, despite its widespread use, YOLO is not the only model suited for object detection in UI analysis. The YOLO family has evolved over time (e.g., YOLOv3, YOLOv5, YOLOv7), with newer versions improving detection accuracy, speed, and generalization to varied input domains. Furthermore, alternative models such as DEtection TRansformer (DETR) [27] represent a significant shift in object detection paradigms by using transformer-based architectures that capture global context directly via self-attention mechanisms. DETR and its successors (like Deformable DETR) have demonstrated competitive performance on benchmark datasets and may offer advantages in UI analysis scenarios involving complex layouts or overlapping components.

Despite their promise, transformer-based models like DETR typically require larger datasets and computational resources, which may present challenges in real-world industrial automation settings. In contrast, one-stage detectors like YOLO remain popular for their balance of speed and accuracy, particularly when integrated into lightweight and scalable test automation pipelines.

In the domain of visual change detection itself, relatively few studies have targeted UI change detection in the context of software test automation. Existing approaches primarily focus on raw pixel or region comparison, sometimes enhanced with simple feature matching, but rarely incorporate structural layout or contextual similarity. Our method addresses this gap by introducing graph-based modeling that explicitly encodes spatial relationships and combines visual, textual, and neighborhood features to assess UI changes robustly. Several related approaches from both UI-specific and broader computer vision research are further discussed in Section 4.6 (Comparative study), where we analyze their similarities and differences with our proposed method in greater detail.

Graph-based representations have been applied in various image processing tasks, including image registration and structural similarity assessment [28-31], but these works largely operate on natural or medical imagery and focus on low-level image features. In contrast, our approach targets the semantic structure of software UIs, identifying meaningful changes that affect usability or functionality. To the best of our knowledge, no prior work in visual test automation has combined deep UI control detection with graph-based semantic modeling in a scalable and domain-specific framework. By leveraging strengths of CNN-based object detectors (YOLO), and integrating structural reasoning, our approach provides a novel and practical solution for robust visual regression testing in software development.

## 3. Graph-based visual change detection method

Our visual change detection model consists of four main steps, i.e. 1) control detection, 2) graph building, 3) graph matching (or graph comparison), 4) and change detection and visualization. We give a detailed description of every step in the following subsections. The proposed visual change detection method operates through a four-stage pipeline, illustrated in Figure 1. First, we apply a deep learning object detection model to identify UI controls in the input screenshots. Second, the detected controls are used to build graph representations of the UI layouts, capturing both spatial and contextual relationships between controls. Third, we match the graphs using a recursive similarity algorithm that compares nodes based on visual, textual, and neighborhood features. Finally, unmatched elements are interpreted as UI changes, which are visualized as heatmaps. This architecture enables the detection of both subtle and structural UI changes in a robust and layout-aware manner.

Given a target image $I$ and its changed version $I'$, the problem is to find coordinates within the images where a change has happened and there is a difference between $I$ and $I'$ in those locations. The detected changes are finally visualized using heatmaps $H$ and $H'$, which show where the changes happen within images $I$ and $I'$.

### 3.1. Control detection

In the first step, user interface controls are detected within the images, i.e. the target and changed images. For this purpose, we utilize an object detection machine learning model, i.e. YOLOv5 [15], specifically trained and tested on a user interface control detection dataset containing 14,155 training and 2,000 test samples we created from software screenshots. Every annotated object belongs to one of twenty four classes specifying the type of user interface controls. The dataset and other experimental details are available at https://github.com/mmoradi-iut/VisualChangeDetection. YOLOv5 is a one-stage object detector; it detects regions of interest and selects the best candidates containing instances of an object of interest end-to-end in one stage. Although the performance of the YOLOv5 object detection model was already discussed on the user interface control detection task in previous work [32], we report the accuracy scores obtained by the model in Table 1, since the performance of the object detector can directly affect the output of the whole image registration method.

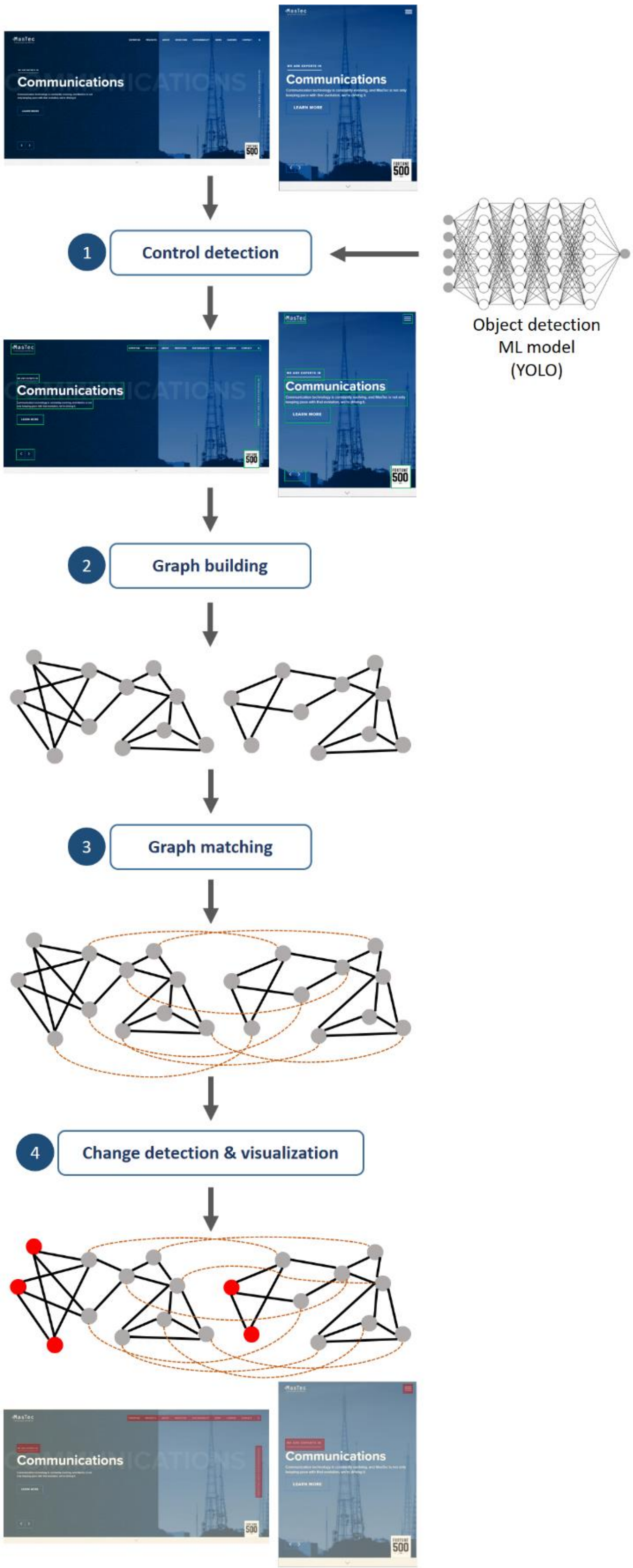

**Figure 1.** The overall architecture of our graph-based change detection method.

**Table 1.** Performance scores obtained by the YOLO object detection model on the user interface control detection test set. The YOLO model is used in the first step of our image registration method to detect user interface controls that appear in the software application screenshot images.

| **Class** | **Precision** | **Recall** | **mAP@.5** | **mAP@.95** |
|---|---|---|---|---|
| ALL | 0.844 | 0.732 | 0.735 | 0.559 |
| ICON | 0.948 | 0.877 | 0.915 | 0.63 |
| DROPDOWN | 0.884 | 0.848 | 0.878 | 0.681 |
| BUTTON | 0.888 | 0.9 | 0.917 | 0.777 |
| MENU | 0.876 | 0.843 | 0.863 | 0.655 |
| INPUT | 0.909 | 0.944 | 0.955 | 0.797 |
| LIST | 0.754 | 0.679 | 0.702 | 0.517 |
| TABBAR | 0.827 | 0.94 | 0.924 | 0.768 |
| TABLE | 0.823 | 0.938 | 0.916 | 0.782 |
| RADIO_SELECTED | 0.923 | 0.87 | 0.919 | 0.644 |
| RADIO_UNSELECTED | 0.853 | 0.919 | 0.873 | 0.642 |
| CHECKBOX_UNCHECKED | 0.919 | 0.856 | 0.857 | 0.592 |
| CHECKBOX_CHECKED | 0.54 | 0.789 | 0.659 | 0.42 |
| TREE | 0.771 | 0.816 | 0.816 | 0.639 |
| IMAGE | 0.926 | 0.94 | 0.934 | 0.794 |
| TEXT | 0.93 | 0.883 | 0.932 | 0.672 |
| LABEL_OF_TEXT_AREA | 0.938 | 0.786 | 0.88 | 0.632 |
| DESCRIPTION_LIST | 0.817 | 0.769 | 0.785 | 0.645 |
| LEGEND | 0.904 | 0.694 | 0.773 | 0.572 |
| HORIZONTAL_AXIS | 0.747 | 0.784 | 0.81 | 0.54 |
| CHART | 0.911 | 0.938 | 0.922 | 0.793 |
| PLOT_TITLE | 0.777 | 0.944 | 0.859 | 0.597 |
| GRAPH | 0.766 | 0.9 | 0.876 | 0.671 |
| VERTICAL_AXIS | 0.631 | 0.311 | 0.309 | 0.218 |
| DATE_AREA | 0.532 | 0.588 | 0.572 | 0.415 |

The inputs of this step are the target and changed images $I$ and $I'$, and the outputs are a set of controls $C=\{c_1, c_2, \ldots, c_N\}$ detected within image $I$, another set of controls $C'=\{c'_1, c'_2, \ldots, c'_M\}$ detected within image $I'$, a set of labels $L=\{l_1, l_2, \ldots, l_N\}$ such that $l_n$ is the class label assigned to control $c_n$, and another set of labels $L'=\{l'_1, l'_2, \ldots, l'_M\}$ such that $l'_m$ is the class label assigned to control $c'_m$.

## 3.2. Graph building

In the second step, an unweighted, undirected graph is built to represent the relationship between the user interface controls within each one of the images. User interface controls detected in the previous step form nodes of the graph. In order to draw edges between the nodes, the Euclidean distance is computed for every pair of nodes, then edges are formed to connect every node to its top $K$ nearest neighbors within the image. This method of constructing a graph from UI control detections based on spatial proximity and contextual layout is, to the best of our knowledge, a novel contribution. It was fully conceived and developed by the authors specifically for the purpose of robust, context-aware visual change detection in software UIs.

Given a set of controls $C=\{c_1, c_2, \ldots, c_N\}$ such that every control is represented as coordinates $(x_1, y_1, x_2, y_2)$ in a two-dimensional space, a list of top $K$ nearest neighbors $NG_n=\{c_1, c_2, \ldots, c_K\}$ is created for every $c_n$. The distance between two controls $c_n$ and $c_j$ is computed using the Euclidean distance as follows:

$$Distance(c_n, c_j) = \sqrt{(x_1 - x'_1)^2 + (y_1 - y'_1)^2 + (x_2 - x'_2)^2 + (y_2 - y'_2)^2} \tag{1}$$

where $(x_1, y_1, x_2, y_2)$ and $(x'_1, y'_1, x'_2, y'_2)$ represent the coordinates of controls $c_n$ and $c_j$, respectively, within image $I$.

Algorithm 1 gives the pseudo code of the procedure for creating a list of top $K$ nearest neighbors for every control within the image. For every control, first, an empty list of neighbors is created (lines 3-4). Second, the distance between target control $c_n$ and every other control $c_j$ is computed (lines 5-6). Control $c_j$ is added to the list of nearest neighbors $NG_n$ if the size of the list is still smaller than $K$ (lines 7-8), otherwise, $c_j$ is compared to every $c_p \in NG_n$. Control $c_p$ is removed from $NG_n$ and $c_j$ is added to $NG_n$ if $c_j$ is closer than $c_p$ to target control $c_n$ (lines 9-11). Finally, a list of top $K$ nearest neighbors is returned for every control $c_n$ (line 13).

The same process is also performed for every control $c'_m$ in $C'$, in order to create a list of top $K$ nearest neighbors $NG'_m=\{c'_1, c'_2, \ldots, c'_K\}$ for every control within the changed image.

**Algorithm 1.** The procedure for creating a list of top $K$ nearest neighbors for controls detected within an image.

**Input:** list of detected controls $C=\{c_1, c_2, \ldots, c_N\}$ within image $I$, number of top nearest neighbors $K$
**Output:** list of top $K$ nearest neighbor controls $NG_n=\{c_1, c_2, \ldots, c_K\}$ for every control $c_n$
**for** every control $c_n \in C$:
    $NG_n=\{\}$
    **for** every other control $c_j \in C$:
        compute the distance between $c_n$ and $c_j$ using Eq. (1)
        **if** size of $NG_n$ is smaller than $K$:
            **then** add $c_j$ to $NG_n$
        **else**:
            **if** there is any $c_p \in NG_n$ where $Distance(c_n, c_j)$ is greater than $Distance(c_n, c_p)$:
                **then** remove $c_p$ from $NG_n$ and add $c_j$ to $NG_n$
    **end for**
    **return** $NG_n$
**end for**

After finding the nearest neighbors for every control, a graph $G=(V, E)$ is built to represent how controls in image $I$ are structured. Graph $G$ is composed of a set of vertexes $V=\{v_1, v_2, \ldots, v_N\}$ where $v_n$ corresponds to control $c_n$, and a set of edges $E=\{e_1, e_2, \ldots, e_N\}$ where $e_n$ connects a pair of vertexes in $V$. A graph $G'=(V', E')$ is also built to represent how controls in image $I'$ are structured. Graph $G'$ is composed of a set of vertexes $V'=\{v'_1, v'_2, \ldots, v'_M\}$ where $v'_m$ corresponds to control $c'_m$, and a set of edges $E'=\{e'_1, e'_2, \ldots, e'_M\}$ where $e'_m$ connects a pair of vertexes in $V'$.

### 3.3. Graph matching

In this step, graphs $G$ and $G'$ are compared to find differences between the graphs, which eventually results in finding changes in the images. We utilize a recursive algorithm to traverse through graphs $G$ and $G'$ and find correspondence between nodes in the graphs.

The main step in this algorithm is to compare every node in $G$ to every node in $G'$ and compute a similarity score for every pair of nodes that have the same class label assigned in the object detection step. The similarity score is computed between nodes $v_n \epsilon V$ and $v'_m \epsilon V'$ by summing up the similarity scores between all neighbors of $v_n$ and $v'_m$ as follows:

$$Similarity(v_n, v'_m) = \sum_{ng_p \epsilon NG_n \text{ and } ng'_q \epsilon NG'_m} Similarity(ng_p, ng'_q) \quad (2)$$

where $NG_n$ is the set of neighbors of $v_n$, $NG'_m$ is the set of neighbors of $v'_m$, $ng_p$ is a neighbor of $v_n$ such that $ng_p \epsilon NG_n$ and $1 \geq p \geq K$, and $ng'_q$ is a neighbor of $v'_m$ such that $ng'_q \epsilon NG'_m$ and $1 \geq q \geq K$.

When computing the similarity between nodes $v_n$ and $v'_m$, the algorithm recursively traverses through neighbors of the nodes and keeps track of nodes it has already seen and computed a similarity value for them. This process continues until the algorithm sees a node it already traversed through and stored in the list of seen nodes. In this situation, a similarity score is computed between $ng_p$ and $ng'_q$, such that $ng_p$ or $ng'_q$ was already seen in previous recursions, by estimating text similarity and/or hash difference between the nodes. The computed similarity value is then propagated to the previous recursions to compute the final similarity score between $v_n$ and $v'_m$. In this case, the similarity score is computed as follows:

$$Similarity(ng_p, ng'_q) = TextSim\left(ng_p, ng'_q\right) + \frac{1}{HashDiff\left(ng_p, ng'_q\right)} \quad (3)$$

where $TextSim(ng_p, ng'_q)$ is the textual similarity between $ng_p$ and $ng'_q$, and $HashDiff(ng_p, ng'_q)$ is the normalized hash difference between $ng_p$ and $ng'_q$.

When a pair of nodes or their neighbors has already been visited in previous recursions, a direct similarity score is computed based on their textual content and visual appearance. Specifically, the textual similarity is calculated using Optical Character Recognition (OCR)-extracted text from the control regions and a similarity metric, i.e. cosine similarity, between text embeddings, producing a score in the range [0, 1]. For control pairs not classified as textual elements, the textual similarity is set to zero. The hash difference is computed by applying a perceptual hashing algorithm, i.e. average hash, to the cropped control regions in the image, measuring the visual difference between them. This raw hash difference, which is always ≥ 0, is then normalized to a [0, 1] range to prevent high hash differences from disproportionately impacting the score. Since lower hash difference values indicate greater visual similarity, we take the inverse of the normalized hash difference. The final similarity score between nodes is then calculated as the sum of the textual similarity and the inverse normalized hash difference, as expressed in Equation (3). This formulation ensures that both the semantic and visual similarities between UI controls are taken into account in the matching process.

After computing similarity scores between nodes in $G$ and $G'$, the change detection method continues with finding the best match for nodes in the graphs. For every node $v_n$ in $V$, a set of candidate matches $CM_n$ is created, and all nodes in $V'$ whose similarity scores with $v_n$ are larger than threshold *sim_threshold* are added to $CM_n$. Finally, the candidate matches are sorted and a node with the highest similarity score is selected as the best match for $v_n$. The graph matching approach introduced in this section, including the recursive neighbor-based similarity computation and the integration of textual similarity and normalized hash difference into node comparison, is a novel contribution of this work. It was specifically developed by the authors to robustly match UI components between software versions based on both structural and visual context.

Algorithm 2 gives the pseudo code of the procedure for computing similarity scores and finding the best match for nodes in the graphs. First, nodes in graph $G$ and $G'$ are compared, if they have the same class label, a similarity score is computed using the recursive procedure given in function *neighbor_sim* (lines 2-6), and a node in $G'$ is added to the target node's list of candidate matches if the similarity score is larger than the threshold (lines 7-8). Then, candidate matches are sorted and a node with the highest similarity score is selected as the best match of the target node. If two or more target nodes share the same node as their best match, the target node having the highest similarity score with the shared node retain it as the best match, and the other target node(s) select a node with the next highest similarity score from their candidate match lists (lines 12-13). The function *neighbor_sim* takes a pair of nodes, their lists of neighbors, and a list of nodes that were already seen in previous recursions of the function (line 15). If both the nodes were not already seen in previous recursions, they are added to the list of nodes that were already seen, and a similarity score is computed by calling the function again and summing up the similarity values resulted from comparing the neighbors (lines 16-22). If one of the nodes were already seen in previous recursions, the similarity is computed by comparing text similarity and/or hash difference between the corresponding controls (lines 23-24).

## 3.4. Change detection and visualization

After performing the graph matching step, those nodes in $G$ and $G'$ that are similar enough, in terms of similarity between their neighbors (and text similarity and hash difference as well), are matched together. However, some nodes may not be matched with any nodes in the other graph. In this case, controls that correspond to these nodes are detected as changes within the images, because no similar controls with similar neighbors were found.

**Algorithm 2.** The procedure for computing similarity scores and finding the best match for nodes in the graphs.

1: **Input:** set of controls $C$={$c_1$, $c_2$, …, $c_N$} detected within reference image $I$, graph $G$=($V$, $E$) that represents the structure of controls in $C$ within $I$, set of controls $C'$={$c'_1$, $c'_2$, …, $c'_M$} detected within changed image $I'$, graph $G'$=($V'$, $E'$) that represents the structure of controls in $C'$ within $I'$, neighbor similarity threshold *sim_threshold*

2: **Output**: best matches between nodes in $V$ and $V'$

3: **for** every node $v_n \in V$:

4: **for** every node $v'_n \in V'$:

5: **if** $v_n$ and $v'_m$ have the same class label:

6: **then** compute the similarity between $v_n$ and $v'_m$ using function *neighbor_sim*()

7: **if** the similarity is larger than *sim_threshold*:

8: **then** add $v'_m$ to list of candidate matches $CM_n$

9: **end for**

10: sort $CM_n$ and select the node with the highest similarity score as the best match for $v_n$

11: **end for**

12: **if** two or more nodes in $V$ share the same $v'_m$ as the best match:

13: **then** node $v_n$ with the highest similarity score remains as the best match with $v'_m$, and other nodes take a candidate from their $CM_n$ with the next highest similarity score

14: **return** a list containing the best match in $V'$ for every node in $V$

15: **function** *neighbor_sim*($v_n$, $NG_n$, $v'_m$, $NG'_m$, *already_seen_list*):

16: **if** $v_n$ and $v'_m$ are not in *already_seen_list* **then**:

17: *temp_similarity* = 0

18: **for** every $ng_p \in NG_n$ and $ng'_q \in NG'_m$:

19: add $ng_p$ and $ng'_q$ to *already_seen_list*

20: *temp_similarity* = *temp_similarity* + *neighbor_sim*($ng_p$, $NG_p$, $ng'_q$, $NG'_q$, *already_seen_list*) //Equation (2)

21: **end for**

22: **return** *temp_similarity*

23: **else if** $v_n$ or $v'_m$ are in *already_seen_list* **then**:

24: **return** similarity between $v_n$ and $v'_m$ using Equation (3)

The visualization is done using heatmaps $H$ and $H'$, which show where in images $I$ and $I'$ there are differences and something has changed. Heatmaps $H$ and $H'$ are two-dimensional matrices with the same width and height as images $I$ and $I'$, respectively, and zero as the initial value of all the elements. If no matching node is found for node $v_n$, the change detection model considers the corresponding control $c_n$ as a control in image $I$ that has changed and no longer appears in $I'$ as it appears in $I$. Therefore, the value of elements in $H$ in the same location as $c_n$ is set to one. Moreover, if no matching node is found for node $v'_m$, the change detection model considers the corresponding control $c'_m$ as a control in image $I'$ that has changed and no longer appears in $I$ as it appears in $I'$. In this way, the value of elements in $H'$ in the same location as $c'_m$ is set to one. We present examples of detected changes within images from different datasets in Section 4.5.

# 4. Experimental results

We conducted a set of experiments to investigate the performance of our graph-based change detection method on datasets of web applications images. Experimental details, data, and codes are available at https://github.com/mmoradi-iut/VisualChangeDetection.

## 4.1. Datasets

We created three different datasets in order to test the performance of our graph-based change detection method in various scenarios. Two datasets were created for change detection within desktop images in two different scenarios. One dataset were created to detect changes between desktop and mobile images.

### 4.1.1. Desktop screenshots

We collected an initial set of images of 250 websites by taking screenshots of the desktop browser in the maximized mode with a screen size of 1,920 by 1,080 pixels. The first dataset was created by automatically applying eight different types of changes to the initial set of images. A maximum of four changes were applied to every image, with a minimum of one change. The number of changes and the type of changes were automatically selected in a random manner. These are the eight change types:

1) ADD CONTROL: It adds a new control to the image in a location where there is no overlap with existing controls.

2) CHANGE LOCATION: It removes a control from its original location and puts it in a new location within the image where there is no overlap with other controls.

3) CHANGE COLOR: It changes the color of pixels within a control in the image. Those pixels that share the same color take the same new color. In this way, the same texts or shapes that appear in the original control still appear in the changed control but in a different color.

4) DUPLICATE: It creates a copy of a control and puts it in the image in a location where there is no overlap with other controls.

5) REMOVE: It selects a control and removes it from the image. The location where the deleted control appeared in the original image is filled with the color of pixels in the neighborhood of that area.

6) RESIZE SMALLER: it resizes a control and makes it smaller by a ratio between 0.3 and 0.8. The area that was a part of the resized control within the original image and no longer belongs to the resized control within the new image (because the control has been resized and occupies les space) is filled with the color of pixels in the neighborhood.

7) RESIZE LARGER: it resizes a control and makes it larger by a ratio between 0.3 and 0.8 if the resized control has no overlap with other controls within the image.

8) SWAP CONTROLS: it selects two controls and swaps their locations if they have no overlap with other controls in the new locations within the image.

Seven changed versions were generated for every image in the dataset by applying different change methods. Therefore, the first dataset contains a total number of 1,750 image pairs, every pair is represented by an original image and a corresponding changed image. Annotations were also generated to specify the location of changes for testing the accuracy.

#### 4.1.2. Desktop screenshots–cut images

We created the second dataset to increase the difficulty of detecting changes and evaluate the accuracy of the visual change detection method in a more complicated scenario. We again generated changed images by applying the eight change methods we introduced in Section 3.1.1, but we also cut a region of every image and shifted the remaining part to the center. In this way, some user interface controls no longer appear in the changed image. Moreover, controls in the changed image do not appear in the same coordinates as in the target (or original) image.

This dataset simulates those scenarios where the active window of the application under test is resized and some controls may not appear anymore in the screenshot. Furthermore, the user interface controls do not appear in the same coordinates as the original screenshot taken before resizing the window. Detecting changes in this scenario can be very challenging, because pixel by pixel comparison easily fails in finding correspondence between controls in the target and changed images.

We created this dataset by applying a maximum of four changes (introduced in the previous subsection), with a minimum of one, to every original screenshot and then cutting a random area of the image in left, right, top, or bottom. The number of changes per image and the type of changes were selected in a random manner. The location of the area that must be cut was selected randomly from the set (left, right, top, bottom), and the number of pixels to cut was also selected randomly from the set (100, 200, 300, 400, 500). Seven changed versions were generated for every image in the dataset by applying different change methods. Therefore, the second dataset contains a total number of 1,750 images, every pair is represented by an original image and a corresponding changed image. Annotations were also generated to specify the location of changes for testing the accuracy.

#### 4.1.3. Desktop–mobile screenshots

The goal of the third dataset is to evaluate the performance of the visual change detection method in detecting differences between screenshots of an application under test in different platforms. We created this dataset by taking screenshots of 180 pairs of different web applications, each pair consisting of one desktop and one mobile screenshot of the same application with the same Graphical User Interface (GUI). The size of images is 1,920 by 1,080 pixels in desktop and 576 by 832 pixels in mobile platform. We also generated ground-truth annotations for every pair of desktop and mobile screenshots by manually annotating user interface controls and correspondence between the controls in the images.

### 4.2. Evaluation metrics

In order to objectively evaluate the performance of our visual change detection method, we used three evaluation measures, i.e. precision, recall, and F-score. Precision, also known as specificity, is an important metric in assessing the accuracy of ML systems. In the context of visual change detection, precision is defined as the ratio of correctly identified changed UI controls to the total number of controls identified by the ML system as changed. The precision score provides insights into the method's ability to minimize false positives, and is calculated as follows:

$$Precision = \frac{True\ positives}{True\ positives + False\ positives} \tag{4}$$

Recall, also known as sensitivity, is another fundamental metric for evaluating the performance of ML systems. In the context of our work, recall measures the ratio of correctly identified changed UI controls to the total number of ground truth changed controls. A high recall score indicates the method's effectiveness in capturing actual changes. The recall is calculated using the following formula:

$$Recall = \frac{True\ positives}{True\ positives + False\ negatives} \tag{5}$$

The F-score, or F-measure, provides a balance between precision and recall. It is particularly useful when there is a need to weigh both false positives and false negatives. The F-score is the harmonic mean of precision and recall and is computed using the following formula:

$$F\text{-}score = 2 \times \frac{Precision \times Recall}{Precision + Recall} \tag{6}$$

We selected precision, recall, and F-score as evaluation metrics because they comprehensively assess the detection model’s ability to correctly identify changes (recall), avoid false positives (precision), and provide a balanced measure (F-score) which is crucial in real-world automated testing scenarios where both over- and under-detection are problematic.

For each metric, we define three variants based on IOU thresholds. A predicted change is considered a true positive if the overlap with ground truth exceeds 0.75, 0.5, or 0.25, respectively, corresponding to precision@0.75, precision@0.5, and precision@0.25. This enables analysis across strict, moderate, and lenient matching conditions.

The IOU measures the overlap between the predicted changed area and the ground truth changed area within an image. In the context of visual change detection, if $A_p$ is the area of the predicted change, $A_g$ is the area of the ground truth change, and $A_o$ is the area of their overlap, then IOU is calculated as follows:

$$IOU = \frac{A_o}{A_p + A_g - A_o} \tag{7}$$

The IOU ranges from 0 to 1, where a value of 0 indicates no overlap, and a value of 1 indicates perfect overlap between the predicted and ground changed areas.

We report these metrics at three IOU thresholds, i.e. 0.75, 0.5, and 0.25, to reflect varying levels of strictness in change localization. A higher IOU threshold (e.g., 0.75) captures fine-grained alignment with the ground truth, suitable for evaluating high-precision tools. Lower thresholds (e.g., 0.25) are more tolerant and highlight whether the system detects the general area of change, useful for more ambiguous or partially occluded visual changes.

## 4.3. Hyperparameter tuning

As explained in Section 3, our visual change detection method has four hyperparameters that control how the method creates a graph, identifies correspondence between UI controls within images, and detects changes. We designed and conducted a set of experiments to investigate the impact of these hyperparameters on the performance of change detection. In this subsection, we present and discuss the experimental results obtained by testing different values of the hyperparameters on each of the three datasets. We used 70 percent of each dataset for tuning the hyperparameters and the remaining 30 percent for testing the change detection method against the baseline change detectors.

### 4.3.1. Desktop screenshots

Figure 2−a) presents the performance scores obtained by our graph-based change detection method in the hyperparameter tuning experiments for different values of the parameter *K*, which specifies the number of top nearest nodes that form neighbors of a node in the graph.

As the results shows, there is a negligible change in precision when *K* takes different values, but recall considerably changes with different values of this hyperparameter. As can be seen, the method obtained the highest recall and f-score when eight neighbors were used for every node to create the graphs. The recall scores suggest that when few numbers of neighbors (e.g. one, two, or three) are used to create the graph, the change detector may suffer from lack of enough information about the contextual similarity between nodes in the graphs. On the other hand, large numbers of neighbors (e.g. more than eight) may mislead the object detector by providing irrelevant information about contextual similarity between nodes.

Figure 2−b) presents the performance scores obtained by our change detection method in the hyperparameter tuning experiments for different values of the parameter *H*, which specifies the maximum hash difference between controls when finding correspondence between them.

As the results show, precision is not substantially affected by different values of *H*, but recall can considerably decrease when *H* takes higher values. This suggests that higher values of this hyperparameter do not provide the change detection method with enough accurate information about the visual similarity between controls, therefore, the method is misled and cannot detect some changes.

Figure 2−c) presents the performance scores obtained by our change detection method in the hyperparameter tuning experiments for different values of the parameter *TS*, which specifies the minimum text similarity between controls when finding correspondence between them.

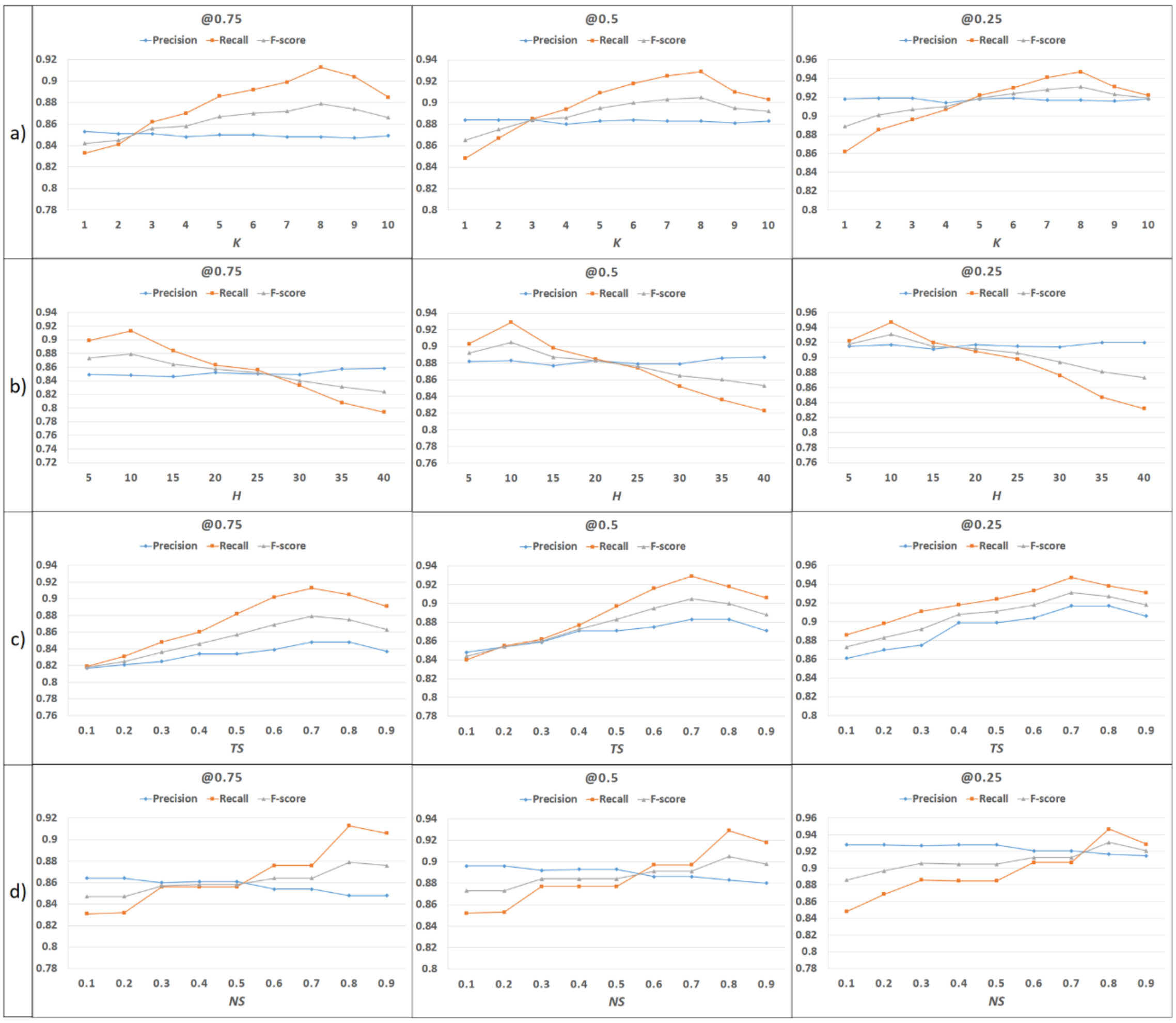

**Figure 2.** Performance scores obtained by our graph-based visual change detection method on the desktop screenshots dataset in the hyperparameter tuning experiments. The results are reported for different values of the parameters a) *K*, which is the number of top nearest neighbors for each node to build the graph, b) *H* that controls the maximum hash difference between UI controls, c) *TS*, which refers to the minimum text similarity between controls that belong to the class TEXT, and d) *NS*, which is the minimum neighbor similarity between nodes in the graph when identifying similar nodes.

As can be seen, both precision and recall are affected by changing the value of *TS*, however, the latter is influenced by a higher degree. The change detector showed its best performance when the minimum textual similarity was set to 0.7. The results demonstrate that smaller textual similarity thresholds mislead the method by allowing dissimilar controls to be identified as corresponding nodes between the graphs. Higher values of *TS* can be also misleading by preventing small degrees of deviation in the textual content of corresponding controls.

Figure 2−d) presents the performance scores obtained by our change detection method in the hyperparameter tuning experiments for different values of the parameter *NS*, which specifies the minimum neighbor similarity between nodes in graphs when finding similar nodes. As reported in Figure 2−d), different values of *NS* can considerably affect the recall scores. The highest scores

were obtained when the minimum neighbor similarity threshold was 0.8, however, smaller values reduced the change detector's performance. This suggests that smaller neighbor similarity thresholds lead to establishing wrong correspondence between controls, because it allows some nodes that do not have enough similar neighbors to be identified as analogous. As a result, the change detection method cannot be able to accurately identify all the corresponding controls, therefore, some changes are not detected.

#### 4.3.2. Desktop screenshots–cut images

In this subsection, we present the hyperparameter tuning results on the desktop screenshots–cut images dataset. Figure 3−a), b), c), and d) present the performance scores obtained by our graph-based change detector for different values of the parameters *K*, *H*, *TS*, and *NS*, respectively.

The results show almost the same patterns we discussed in Section 4.3.1 for the hyperparameter tuning experiments on the desktop screenshots dataset. However, the best results are reported for *K*=6 when experimenting with different values on the desktop screenshots–cut images dataset. The reason may be that some regions of the changed images were cut in this dataset, hence, some controls missed their closer neighbors in the UI. These controls were consequently connected to farther neighbors in the graph due to a larger value of *K*, e.g. eight. Being connected to farther neighbors in the graph created for the changed image can cause the change detection model being misled in finding proper correspondence between controls, since the corresponding control in the original image is still connected to the closer neighbors in the graph. Therefore, a smaller value of *K*, e.g. six, can help the method to perform more accurately in detecting changes by considering fewer neighbors in the graph in scenarios where not all the original UI controls are visible in the changed image (for instance, when the application window has been resized, or some part of it is covered by other windows).

#### 4.3.3. Desktop–mobile screenshots

In this subsection, we present the hyperparameter tuning results on the desktop–mobile screenshots dataset. Figure 4−a), b), c), and d) give the performance scores obtained by our graph-based change detector for different values of the parameters *K*, *H*, *TS*, and *NS*, respectively.

The results show almost the same patterns we discussed in Section 4.3.1 and Section 4.3.2 for the hyperparameter tuning experiments on the other two datasets. However, the best results are reported for *K*=5 when experimenting with different values on the desktop–mobile dataset. The reason can be fewer number of controls and lower control density in the mobile images. When more numbers of neighbors, e.g. eight or nine, are used to create the graph for the mobile image, some neighbors may be too distant from a target node, whereas the same controls may not be connected to the corresponding node in the graph created for the desktop image. As a result, the graph cannot provide the change detector with highly accurate information about the most relevant neighbors of every control. On the other hand, smaller values of the parameter $K$, e.g. five or six, lead to discarding farther neighbors, which eventually help the change detector identify correspondence between controls with respect to their most relevant neighbors.

Based on the tuning experiments across all three datasets, we selected the following final hyperparameter values for the comparative evaluation in Section 4.5:

- Number of nearest neighbors $K$=8 for desktop screenshots, $K$=6 for cut images, and $K$=5 for desktop–mobile pairs;
- Maximum hash difference $H$=10;
- Minimum text similarity threshold $TS$=0.7;
- Minimum neighbor similarity threshold $NS$=0.8.

These values consistently resulted in high F-scores and a good trade-off between precision and recall across varying IOU thresholds. Notably, adjusting $K$ per dataset improved robustness when layout or platform changes altered neighborhood structures. All comparative results reported in Section 4.5 were obtained using these tuned parameters.

## 4.4. Baselines

We designed and conducted another set of experiments to evaluate the performance of our visual change detection method against other change detectors. Given the innovative nature of our approach, there exists a lack of directly comparable methods in the existing literature. Existing visual change detection tools often rely on pixel-level comparisons, lacking the sophistication necessary for nuanced UI change detection. In the absence of directly analogous methods, we implemented two baseline change detectors to establish a meaningful context for evaluating our proposed method. These baselines are designed to represent conventional strategies for change detection in software interfaces and serve as benchmarks against which the performance of our method can be objectively assessed. We describe these two baselines in the following subsections.

### 4.4.1. Pixel-wise comparison

This baseline method involves a pixel-wise comparison of the original and modified software screenshots. Pixels in the same location within the original and changed images are compared against each other and that location is marked as a change if the pixel values are different. This approach, although simplistic, reflects a common strategy employed in many visual change detection methodologies. However, it fails to capture the semantic relationships between user interface controls and has limited adaptability to dynamic interface changes.

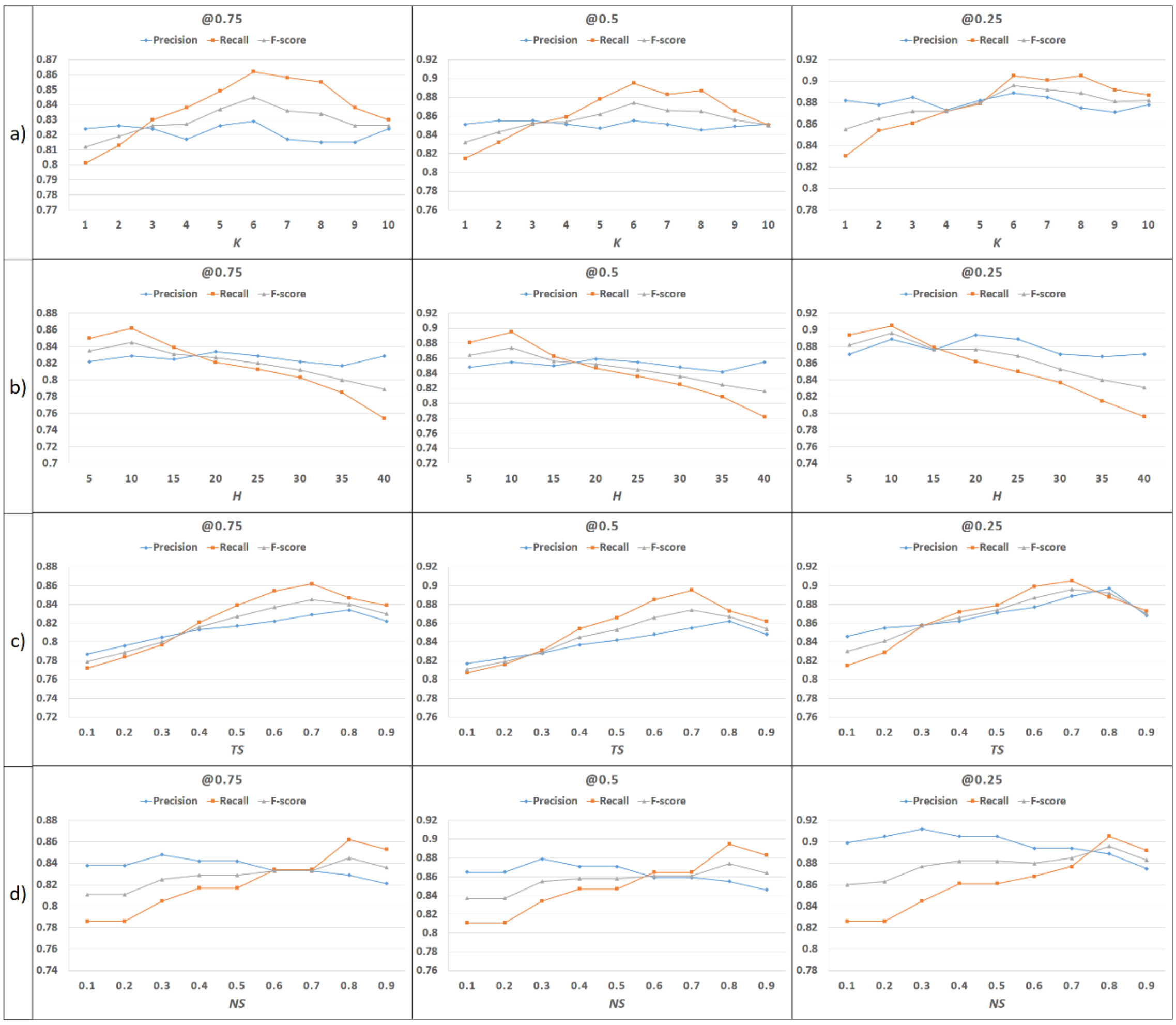

**Figure 3.** Performance scores obtained by our graph-based visual change detection method on the desktop screenshots–cut images dataset in the hyperparameter tuning experiments. The results are reported for different values of the parameters a) *K*, which is the number of top nearest neighbors for each node to build the graph, b) *H* that controls the maximum hash difference between UI controls, c) *TS*, which refers to the minimum text similarity between controls that belong to the class TEXT, and d) *NS*, which is the minimum neighbor similarity between nodes in the graph when identifying similar nodes.

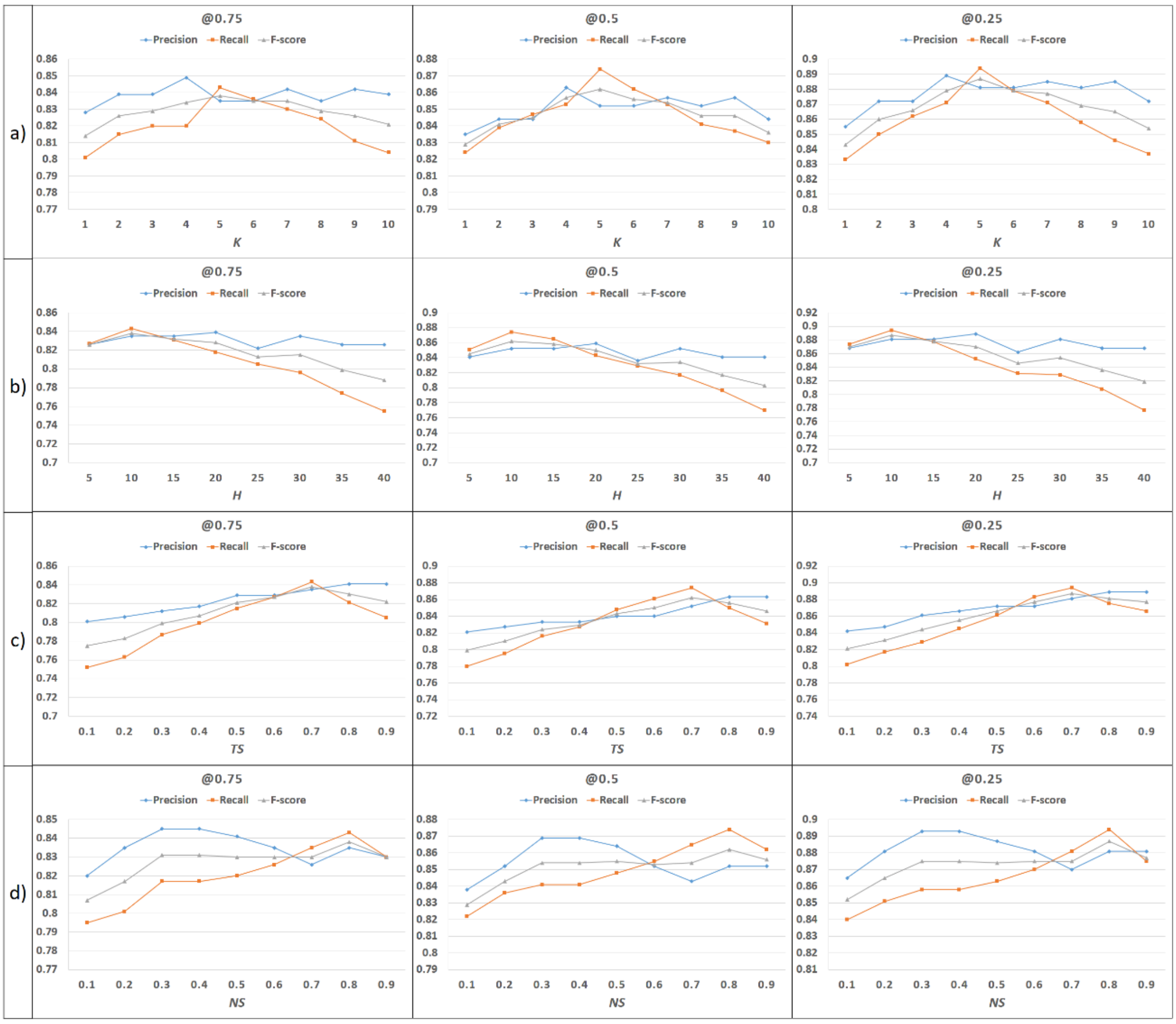

**Figure 4.** Performance scores obtained by our graph-based visual change detection method on the desktop–mobile screenshots dataset in the hyperparameter tuning experiments. The results are reported for different values of the parameters a) *K*, which is the number of top nearest neighbors for each node to build the graph, b) *H* that controls the maximum hash difference between UI controls, c) *TS*, which refers to the minimum text similarity between controls that belong to the class TEXT, and d) *NS*, which is the minimum neighbor similarity between nodes in the graph when identifying similar nodes.

#### 4.4.2. Region-based change detection

The second baseline method employs a region-based approach, wherein predefined regions of interest, i.e. UI controls, within the software screenshot are analyzed for changes. UI controls are first detected using the YOLO object detection model, then, controls in the same location within the images are compared using the hash difference. Controls are marked as a change if they have a hash difference higher than 10 with their corresponding region within the other image. This hash difference threshold was specified by hyperparameter tuning on the training datasets. While this baseline method introduces a level of abstraction beyond pixel-wise comparison, it may struggle with capturing spatial relationships and nuanced changes that extend beyond control-wise visual similarity.

### 4.5. Performance evaluation against baselines

Table 2, 3, and 4 present the performance scores obtained by our graph-based visual change detection method and the two baselines on the desktop screenshots, desktop screenshots–cut images, and desktop–mobile screenshots test datasets, respectively.

As Table 2 shows, the pixel-wise comparison (PWC) baseline can perform better than the other methods in the simplest scenario, i.e. detecting UI changes when the original and changed screenshots come from the same modality (desktop UI in this case), with the same size, no resizing, and no cut regions. However, in more complicated scenarios, as can be seen in Table 3 and 4, the pixel-wise comparison fails to properly detect changes.

Our graph-based change detector could not perform better than the pixel-wise comparison baseline on the desktop screenshots dataset. However, it achieved a precision of more than 85 percent and a recall of more than 92 percent. On the other two datasets, our change detection method outperformed the two baselines.

It is worth noting that, across all experimental evaluations, systematically higher precision, recall, and F-score values were observed at lower IOU thresholds (e.g., 0.25) compared to higher ones (e.g., 0.75). This behavior is expected and aligns with standard practices in evaluating object detection systems. Lower IOU thresholds are more lenient, allowing detections that approximately cover the changed regions to count as true positives, even if they do not align perfectly. This is particularly relevant in visual UI change detection where control bounding boxes may have shifted or been resized between versions, causing slightly misaligned predictions. Therefore, the higher scores at 0.25 reflect the method's capacity to detect general regions of change effectively, even if precise alignment varies slightly.

**Table 2.** Performance scores obtained by our graph-based visual change detection method and the baselines on the desktop screenshots test dataset. The highest score in each column is shown in underlined face. PWC: Pixel-wise Comparison, RCD: Region-based Change Detection, GVCD: our Graph-based Visual Change Detection.

| | **@0.75** | | | **@0.5** | | | **@0.25** | | |
|---|---|---|---|---|---|---|---|---|---|
| | **Precision** | **Recall** | **F-score** | **Precision** | **Recall** | **F-score** | **Precision** | **Recall** | **F-score** |
| PWC baseline | <u>1.000</u> | <u>0.931</u> | <u>0. 964</u> | <u>1.000</u> | <u>0.948</u> | <u>0. 973</u> | <u>1.000</u> | <u>0.961</u> | <u>0. 980</u> |
| RCD baseline | 0.762 | 0.775 | 0. 768 | 0.786 | 0.802 | 0. 793 | 0.803 | 0.819 | 0. 810 |
| GVCD | 0.853 | 0.922 | 0. 886 | 0.891 | 0.935 | 0. 912 | 0.927 | 0.953 | 0. 939 |

**Table 3.** Performance scores obtained by our graph-based visual change detection method and the baselines on the desktop screenshots–cut images test dataset. The highest score in each column is shown in underlined face. PWC: Pixel-wise Comparison, RCD: Region-based Change Detection, GVCD: our Graph-based Visual Change Detection.

| | **@0.75** | | | **@0.5** | | | **@0.25** | | |
|---|---|---|---|---|---|---|---|---|---|
| | **Precision** | **Recall** | **F-score** | **Precision** | **Recall** | **F-score** | **Precision** | **Recall** | **F-score** |
| PWC baseline | - | - | - | - | - | - | - | - | - |
| RCD baseline | 0.656 | 0.684 | 0. 669 | 0.692 | 0.704 | 0. 697 | 0.712 | 0.719 | 0. 715 |
| GVCD | <u>0.835</u> | <u>0.870</u> | <u>0. 852</u> | <u>0.862</u> | <u>0.899</u> | <u>0. 880</u> | <u>0.897</u> | <u>0.911</u> | <u>0. 903</u> |

**Table 4.** Performance scores obtained by our graph-based visual change detection method and the baselines on the desktop–mobile screenshots test dataset. The highest score in each column is shown in underlined face. PWC: Pixel-wise Comparison, RCD: Region-based Change Detection, GVCD: our Graph-based Visual Change Detection.

| | **@0.75** | | | **@0.5** | | | **@0.25** | | |
|---|---|---|---|---|---|---|---|---|---|
| | **Precision** | **Recall** | **F-score** | **Precision** | **Recall** | **F-score** | **Precision** | **Recall** | **F-score** |
| PWC baseline | - | - | - | - | - | - | - | - | - |
| RCD baseline | 0.620 | 0.632 | 0. 625 | 0.651 | 0.649 | 0. 650 | 0.673 | 0.664 | 0. 668 |
| GVCD | <u>0.841</u> | <u>0.849</u> | <u>0. 844</u> | <u>0.862</u> | <u>0.880</u> | <u>0. 871</u> | <u>0.887</u> | <u>0.905</u> | <u>0. 896</u> |

Figure 5 shows a pair of images from the desktop screenshots dataset, and heatmaps that show visual changes detected by our graph-based change detection method. As can be seen, all the changes have been accurately detected within this pair of screenshots. The pixel-wise comparison can also perfectly perform on this dataset since the screenshot pairs were exactly taken from the same location in the applications, without resizing the window and cutting some parts of the images. In this simple scenario, it is imperative for change detection methods to exhibit a high degree of precision and recall.

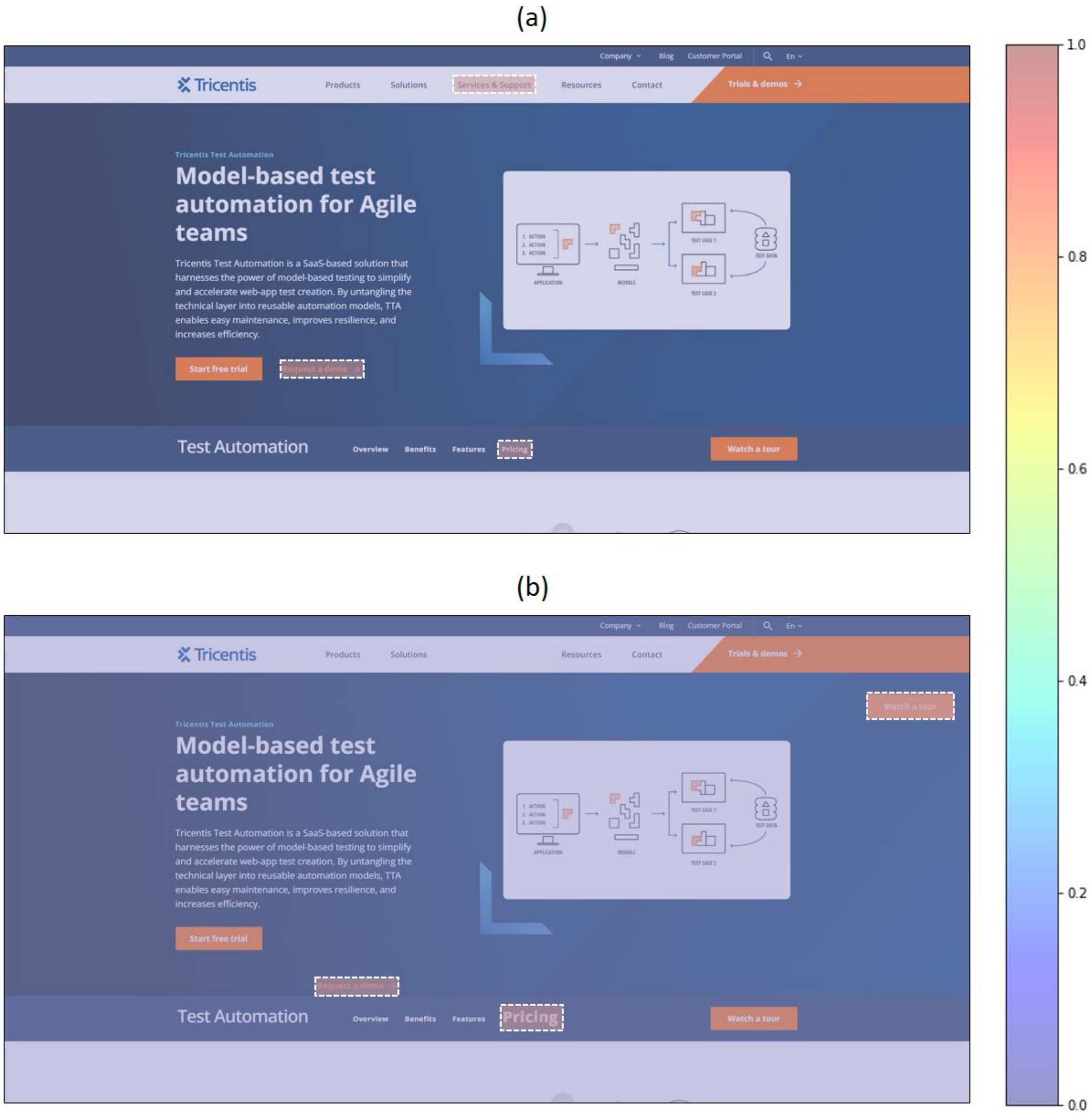

**Figure 5.** A pair of images from the desktop screenshots dataset, and heatmaps that show visual changes detected by our graph-based change detection method. (a) The original screenshot, (b) the changed screenshot. The changes are shown by rectangles with white dashed-line borders.

Figure 6 shows a pair of images from the desktop screenshots–cut images dataset, and heatmaps that show visual changes detected by our graph-based change detection method. As the images show, in this scenario, some part of the image is not visible in the changed version (this simulates resizing the software/application window). Therefore, the UI controls do not appear in the same coordinate within the changed image as the original screenshot. In this case, the pixel-wise comparison totally fails to detect changes because it relies on comparing pixels in the same coordinate between the images. The region-based baseline performs better than the pixel-wise comparison, however, it underperforms the graph-based change detector. The graph model provides the change detector with a context-aware representation of the UI layout that can effectively identify which UI element deviates from its original context within the target image.

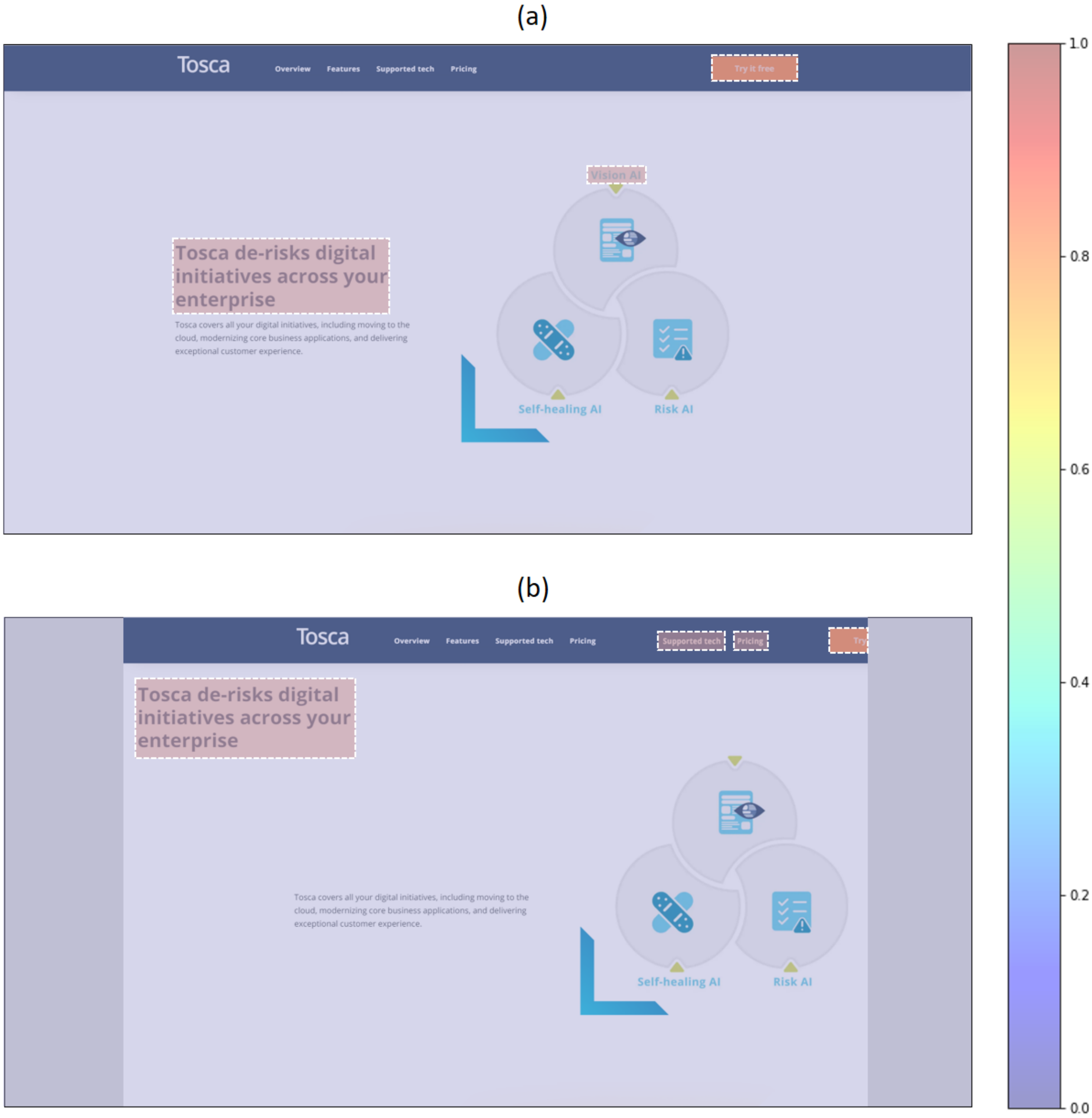

**Figure 6.** A pair of images from the desktop screenshots–cut images dataset, and heatmaps that show visual changes detected by our graph-based change detection method. (a) The original screenshot, (b) the changed screenshot. The changes are shown by rectangles with white dashed-line borders.

Figure 7 shows a pair of images from the desktop–mobile screenshots dataset, and heatmaps that show visual changes detected by our graph-based change detection method. As can be seen, our method has been also able to accurately detect differences between the screenshots of the same application from desktop and mobile platforms. The pixel-wise comparison totally fails in this scenario as well. The region-based baseline performed better than the other baseline, however it still underperforms the graph-based change detector since it suffers from lack of contextual information. These examples of various visual testing scenarios demonstrate the effectiveness of our change detection model for identifying visual regressions in automated software testing and quality assurance.

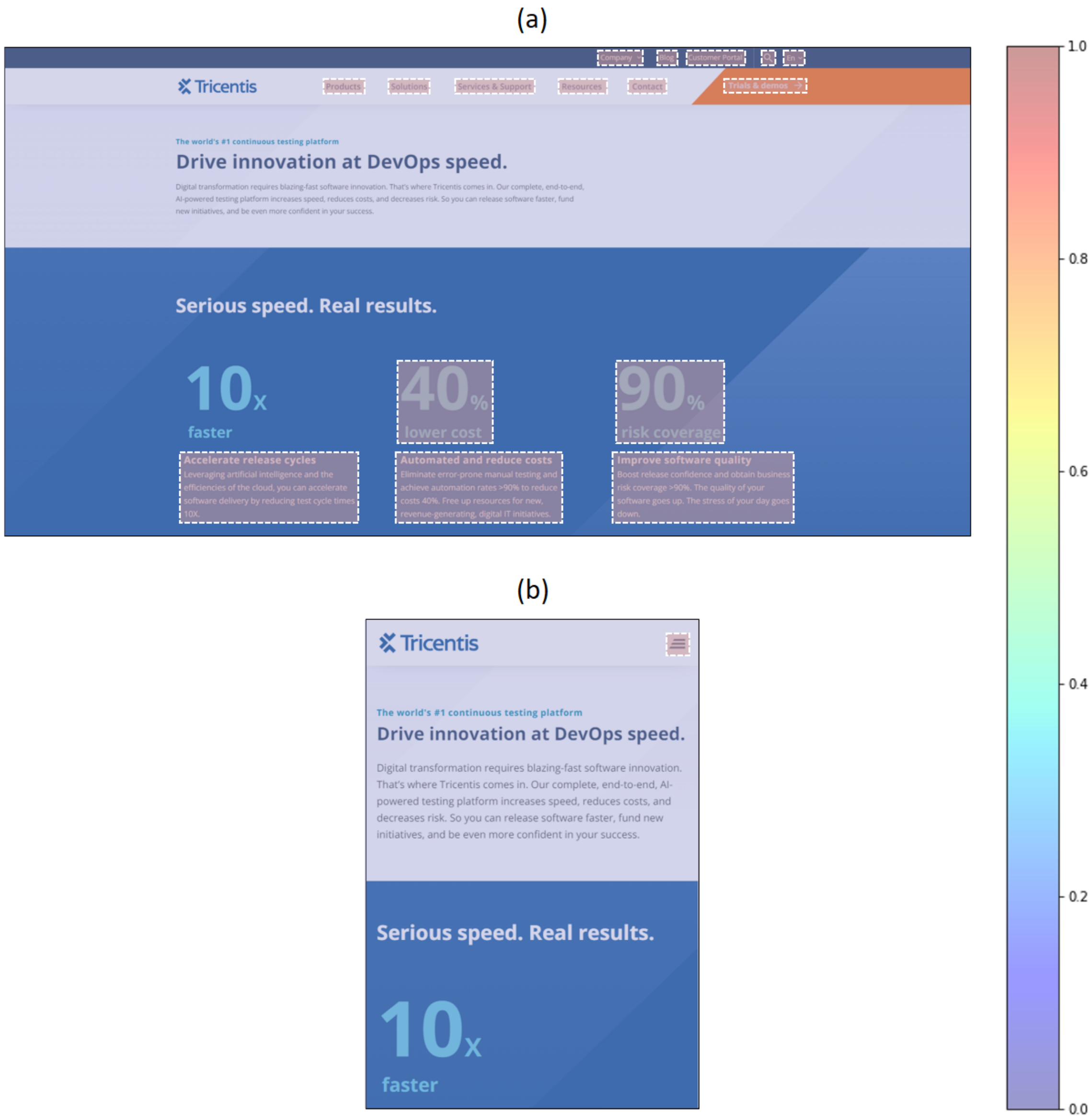

**Figure 7.** A pair of images from the desktop–mobile screenshots dataset, and heatmaps that show visual differences detected by our graph-based change detection method. (a) The original screenshot from desktop platform, (b) the screenshot from mobile platform. The differences are shown by rectangles with white dashed-line borders.

## 4.6. Comparative discussion

In addition to the baseline evaluations presented above, we provide a comparative discussion of our approach against conceptually related methods from both UI-specific and general visual change detection literature.

A method most similar to our UI change detector is Detecting Outdated Screenshots in Documentation (DOSUD) [33], which employs deep learning to determine whether a UI screenshot in software documentation is outdated. DOSUD automatically extracts screenshots from documents and trains a classifier to recognize subtle UI changes in specific regions of the image. However, this approach reduces the problem to an image-level classification task and does not attempt to localize the precise changes within the UI. As such, DOSUD cannot be directly evaluated on the datasets used in our experiments, which require fine-grained change localization. In contrast, our method operates at the UI control level, enabling accurate detection and contextual interpretation of changes within the interface.

In the broader computer vision field, several change detection networks have been proposed for domains outside UI analysis. ChangeNet [34] is one notable example, a deep two-stream CNN architecture designed to localize and label changes between image pairs in street view and aerial imagery. While effective in scene understanding scenarios, ChangeNet is trained on natural images and cannot be directly applied to UI change detection datasets without substantial retraining. Moreover, its large CNN backbone would require a very large corpus of labeled software screenshots to achieve comparable performance in our domain, which is challenging in practice. In contrast, our approach requires training only the UI control detection component on image data, while the change detection stage operates on a graph representation of the UI and does not require extensive data for retraining.

Another influential approach in broader vision research uses Siamese neural networks for change segmentation. For instance, Yang et al. [35] treat change detection as a binary image segmentation problem, producing a mask of changed areas via a Siamese U-Net. While this method performs well in remote sensing tasks involving aerial images of the same location, it assumes spatial alignment and consistent scene framing between the compared images. This assumption does not hold in many UI change detection scenarios, where screenshots may differ significantly in layout, resolution, or platform (e.g., web versus mobile). In such cases, segmentation-based methods may misinterpret layout variations as entirely different scenes. Our graph-based, context-aware approach is robust to these differences, allowing detection of meaningful changes even when overall image framing or layout varies substantially.

### 4.7. Limitations

One of the limitations of this study is the use of relatively simple baseline methods, pixel-wise comparison and region-based change detection, for performance evaluation. While these baselines are widely used and serve as common strategies in visual UI change detection, they do not fully represent the breadth of existing methods in the broader visual change detection literature, particularly those utilizing more complex models such as deep semantic segmentation or transformer-based architectures. However, our primary motivation for selecting these baselines was the lack of directly comparable prior work focused on context-aware UI change detection. As such, we aimed to establish a clear and interpretable benchmark to demonstrate the effectiveness of our method in a real-world test automation setting. For future work, we plan to implement and evaluate more advanced baseline methods, including state-of-the-art visual comparison techniques, to provide a broader performance context and further validate the robustness of our proposed approach.

Another limitation of our current study is that we did not compare the performance of different object detection models used for UI control detection. While we employed YOLOv5 due to its strong trade-off between speed and accuracy, we acknowledge that other models, particularly transformer-based architectures such as DETR, could offer complementary benefits. DETR, for instance, captures global context using self-attention mechanisms and may perform better in detecting overlapping or complex UI layouts. However, due to time and computational resource constraints, we were not able to include DETR or similar models in our experimental evaluation. Exploring the effectiveness of DETR and other state-of-the-art object detectors in the context of UI control detection is a promising direction for future work.

Another limitation of our study lies in the dataset size and the range of test scenarios. While we aimed to differentiate between simple and complex UI change detection cases, the number of samples in each category may not be sufficiently large to fully support a robust distinction between simple and complex change scenarios. This limitation stems from the manual process of collecting and curating high-quality software screenshots to ensure clean and well-labeled data. In future work, we plan to significantly expand the dataset, automate parts of the data collection process, and include a broader spectrum of test cases covering a wider range of simple to highly complex UI modifications.

## 5. Conclusion

In this paper, we presented a novel method for visual change detection in software test automation, combining computer vision, ML, and graph-based analysis. Our approach, which involves detecting UI controls, constructing context-aware graph models, and comparing correspondences between the original and changed versions of software screenshots, addresses the limitations of traditional pixel-wise and region-based techniques. Through extensive experimentation, we demonstrated the method's efficacy in detecting various interface changes and its superiority over baseline methods in more complex visual testing scenarios. The use of a ML model for control detection and the subsequent graph-based analysis not only enhances the precision of change detection but also provides a contextual understanding of the spatial relationships between controls, as well as their visual and textual content. This holistic representation of the UI contributes to the robustness of our method, allowing it to adapt to diverse simple and complex testing scenarios.

The experimental results underscore the practical significance of our method in enhancing the accuracy, efficiency, and reliability of visual change detection, which plays an integral role in software test automation. As software systems evolve and interface intricacies grow, the presented approach offers a forward-thinking solution to address the challenges associated with dynamic user interfaces. Looking ahead, our work opens avenues for further exploration at the intersection of AI, and software testing methodologies. The proposed method not only contributes to the advancement of automated software testing but also holds promise for future research endeavors, fostering innovation in the dynamic landscape of software development and quality assurance.

Building on the foundational work of utilizing AI for detecting visual changes in software screenshots, future research could focus on several promising avenues. One potential area is the integration of DL techniques to enhance the granularity and accuracy of change detection, enabling the system to identify subtle differences that might affect UX but are currently undetectable with existing methods. Additionally, the development of AI-driven predictive models could foresee possible UI inconsistencies caused by changes in the software environment or updates, allowing developers to proactively adjust designs. Another area for expansion is the application of reinforcement learning to continuously improve test automation processes based on feedback loops from past testing cycles. This could significantly reduce human intervention in UI testing, leading to more efficient and less error-prone software development pipelines. Finally, exploring the intersection of AI with augmented and virtual reality could offer innovative ways to test and ensure the quality of immersive user interfaces, catering to the evolving demands of modern software applications.

## Statements and Declarations

The authors have no relevant financial or non-financial interests to disclose. The authors have no conflicts of interest to declare that are relevant to the content of this article.